\documentclass
[aps,
twocolumn,
superscriptaddress,
amsmath,
amssymb
]
{revtex4}

	\usepackage{graphicx}
	\usepackage{dcolumn}
	\usepackage{bm}
	\usepackage{color}
	\usepackage{url}
	\usepackage{tabularx}
	\usepackage{array}
	\usepackage{booktabs}
	\usepackage{comment}
	\usepackage{footnote}
	\usepackage[normalem]{ulem}
	\usepackage{longtable}
	\usepackage{multirow}
	
	\usepackage[T1]{fontenc}
	\usepackage[utf8]{inputenc}
	\usepackage[english]{babel}
	
	\usepackage[cmyk,dvipsnames]{xcolor}
	
	\usepackage{braket}
	
    \definecolor{pacificb}{HTML}{1CA9C9}
    
    \newcommand{\encircle}[1]{\protect\raisebox{.5pt}{\textcircled{\protect\raisebox{-.9pt} {{#1}}}}}

    \newcolumntype{Y}{>{\centering\arraybackslash}X}

\setcitestyle{super}

\begin{document}

\title{Lifetime of racetrack skyrmions}


\newcommand{\RWTH}{Department of theoretical solid state Physics, Rheinisch Westfälische Technische Hochschule Aachen}
\newcommand{\Juelich}{Peter Gr\"unberg Institut and  Institute for Advanced Simulation,  Forschungszentrum J\"ulich and JARA, D-52425  J\"ulich, Germany}
\newcommand{\KTHStockholm}{KTH Royal Institute of Technology, Stockholm SE-10691, Sweden}
\newcommand{\KTH}{Department of Applied Physics, School of Engineering Sciences, KTH Royal Institute of Technology, Electrum 229, SE-16440 Kista, Sweden}
\newcommand{\KTHTwo}{SeRC (Swedish e-Science Research Center), KTH Royal Institute of Technology, SE-10044 Stockholm, Sweden}
\newcommand{\Uppsala}{Department of Physics and Astronomy, Uppsala University, Box 516, SE-75120 Uppsala, Sweden}
\newcommand{\SPSU}{Department of Physics, St. Petersburg State University, St. Petersburg, 198504, Russia}
\newcommand{\HI}{Science Institute of the University of Iceland, 107 Reykjavik, Iceland}
\newcommand{\ITMO}{ITMO University, 197101 St. Petersburg, Russia}
\newcommand{\Aalto}{Aalto University, FI-00076 Espoo, Finland}

\author{Pavel F. Bessarab}%
    \email{bessarab@hi.is}
    \affiliation{\HI}
    \affiliation{\ITMO}

\author{Gideon P. M\"uller}
    \affiliation{\HI}
    \affiliation{\Juelich}
    
\author{Igor S. Lobanov}
    \affiliation{\ITMO}

\author{Filipp N. Rybakov}
    \affiliation{\KTHStockholm}
    
\author{Nikolai S. Kiselev}
	\affiliation{\Juelich}

\author{Hannes J\'onsson}
	\affiliation{\HI}
    \affiliation{\Aalto}
	
\author{Valery M. Uzdin}
	\affiliation{\SPSU}
	\affiliation{\ITMO}

\author{Stefan Bl\"ugel}
	\affiliation{\Juelich}

\author{Lars Bergqvist}
    \affiliation{\KTH}
    \affiliation{\KTHTwo}

\author{Anna Delin}
    \email{annadel@kth.se}
    \affiliation{\KTH}%
    \affiliation{\KTHTwo}
    \affiliation{\Uppsala}

\date{\today}

\begin{abstract}
The skyrmion racetrack is a promising concept for future information technology. There, binary
bits are carried by nanoscale spin swirls -- skyrmions -- driven along magnetic strips. Stability of the skyrmions is a critical issue for the realization of this technology. 
Here we demonstrate that the racetrack skyrmion lifetime can be calculated from first principles as a function of temperature, magnetic field and track width. Our method combines harmonic transition state theory extended to include Goldstone modes, with an atomistic spin Hamiltonian parametrized from density functional theory calculations.
We demonstrate that two annihilation mechanisms contribute to the skyrmion stability: At low external magnetic field, escape through the track boundary prevails, but a crossover field exists, above which the collapse in the interior becomes dominant. Considering a Pd/Fe bilayer on an Ir(111) substrate as a well-established model system, the calculated lifetime is found to be consistent with reported experimental measurements. Our results open the door for predictive simulations, free from empirical parameters, to aid the design of skyrmion-based information technology.
\end{abstract}

\maketitle

\section{Introduction}

Spin textures with topological charge, also called skyrmions~\cite{Bogdanov1989,Bogdanov1994}, hold great promise as a basis for a new type of information technology~\cite{kiselev_2011,fert_2013,mueller_2017}. In particular, information flow can be associated with metastable skyrmions driven along a magnetic strip, as suggested in skyrmion racetrack schemes~\cite{fert_2013,mueller_2017}. It has been demonstrated that skyrmions are sensitive to controlled external stimuli such as electric current~\cite{yu_2012,sampaio_2013,woo_2016}, which is beneficial for efficient, low power data processing. For such a technology to be viable, however, the skyrmion lifetime, $\tau$, is an essential quantity. It is a quantitative measure of stability and needs to be long enough to enable information storage with negligible loss. Prediction of the lifetime of skyrmions 
in arbitrary materials and materials combinations as a function of temperature and various external parameters such as applied magnetic field is thus a necessary prerequisite for developing an optimized skyrmion-based technology. Although skyrmions owe their stability to topology, the lifetime cannot be derived from topological considerations {\it per se}. The celebrated notion of topological protection of single skyrmions localised in a ferromagnetic ground state of infinite spatial dimensions described in the language of continuum theory translates to finite energy barriers, whose heights are described in practise by the escape of the skyrmion into the ferromagnetic state through a Bloch point, or the presence of a discrete lattice of atoms or finite size boundaries. 

Here, we show that it is indeed possible to calculate -- from first principles -- the lifetime of skyrmions. To demonstrate our method, we present results for a  fcc stacked film of one monolayer of Pd and Fe on an Ir(111) substrate, one of the best investigated systems hosting single N\'eel-type skyrmions stabilized by interface generated Dzyaloshinskii-Moriya (DM) interaction~\cite{moriya_1960,dzyaloshinskii_1957}.  We compare the results to an hcp stacked Pd film on Fe/Ir(111), which emerges experimentally as a structurally metastable state~\cite{kubetzka_2017}.
It is experimentally known that at low temperatures this system exhibits a spin-spiral ground state with a period of 6-7~nm at zero applied magnetic field. As the magnetic field is increased, skyrmions, a few nanometers in diameter, are observed to coexist with the spiral ordering, followed by a pure hexagonal skyrmion lattice and finally, at a field of about 2~T, a field-polarized ferromagnetic phase~\cite{romming_2013} with single skyrmions stable above fields of 2.5~T. This sequence of phases can be reproduced by an atomistic spin Hamiltonian (see Methods section) parametrized from 
first principles density functional theory (DFT) calculations~\cite{dupe_2014,simon_2014,rozsa_2016,malottki_2017}. The calculations predict that the ferromagnetic phase emerges at 0.5~T and 3.2~T for hcp and fcc stacking of the Pd layer, respectively, while the skyrmion size ranges from 2 to 3~nm at 4~T, depending on the Pd layer stacking~\cite{malottki_2017}. These theoretical results are in good agreement with the experimental measurements~\cite{romming_2013,kubetzka_2017}. A critical temperature of the phase transition to the paramagnetic state was calculated to be between 227~K~\cite{boettcher_2017} and 250~K~\cite{rozsa_2016}, for independently determined microscopic parameter sets and the temperatures are independent of the magnetic field strength~\cite{boettcher_2017}. However, the temperature range at which skyrmions in this system become stable on macroscopic time scales remains unexplored, although some rough estimates obtained by extrapolation of Monte Carlo and spin dynamics simulations to low temperatures exist~\cite{hagemeister_2015,rozsa_2016}.

Our approach to the identification of skyrmion lifetimes combines an atomistic Hamiltonian parametrized from first principles and statistical rate theory, which provides a solution to the rare-event problem.  The rare-event problem arises from the fact that in the relevant temperature range, transitions between stable magnetic states, e.g. skyrmion collapse to the ferromagnetic phase, induced by thermal fluctuations are typically rare events on the intrinsic time scale of the magnetization dynamics of the system and makes direct simulations of finite temperature spin dynamics~\cite{rozsa_2016,rohart_2016} an intractable approach to evaluate the lifetimes. 

The rate theories~\cite{vineyard_1957,kramers_1940,langer_1969,bessarab_2012} predict an Arrhenius expression for the transition rate $k$ as a function of temperature $T$,
\begin{equation}
    k (T)= \frac{1}{\tau (T)} = \nu(T) e^{-\Delta E/k_\text{B}T},
\label{eq:lifetime}
\end{equation}
where the magnitudes of both the activation energy $\Delta E$ and the entropic pre-exponential factor also referred to as the attempt frequency $\nu$ depend on the parameters of the system as well as the mechanism of the magnetic transition.

Previously, the
mechanism and energy barrier of skyrmion annihilation in magnetic monolayers have been identified~\cite{bessarab_2015,lobanov_2016}. 
The mechanism involves symmetrical shrinking and collapse of the skyrmion via 
an intermediate state reminiscent of the cross section texture of a Bloch point. 
In finite-size systems, the boundaries may provide additional paths for the skyrmion creation and annihilation. 
Quantitative assessment of the effect of boundaries is particularly important in the context of the racetrack memory device, where skyrmions need to be guided reliably along a magnetic strip~\cite{fert_2013}. While the repulsive 
interaction between skyrmions and edges has 
been pointed out before~\cite{rohart_2013,meynell_2014}, and corresponding energy barriers have been evaluated~\cite{stosic_2017,uzdin_2017,cortes_2017}, the impact of the system boundaries on the skyrmion lifetime at a finite temperature is lacking. 

Most previous work on skyrmion stability 
has relied on an 
effective, nearest-neighbor approximation for the interatomic magnetic exchange interaction entering the atomistic spin Hamiltonian. This approach is sufficient for the description of equilibrium properties of skyrmions such as size and shape as well as zero-temperature phase diagrams. 
However, a recent study of skyrmion stability in Pd/Fe/Ir(111) system~\cite{malottki_2017} employing {\it ab initio} calculations of shell-resolved exchange interaction parameters and identification of minimum energy paths for the skyrmion annihilation has demonstrated that a nearest-neighbor description of magnetic interactions greatly underestimated the energy barrier for skyrmion collapse. Therefore, including terms beyond nearest-neighbor pairwise interaction is critical for the quantitative description of the skyrmion stability in itinerant electron magnets, for which long-range, frustrated exchange is a typical feature. 

The evaluation of the attempt frequency $\nu$ is essential for the calculation of the skyrmion lifetime. While the activation energy defines strong, exponential dependence of the lifetime on temperature, it is the attempt frequency 
that establishes the timescale. It can vary by several orders of magnitude depending on parameters of the magnetic system, as has been demonstrated both experimentally and theoretically for 
Fe islands on W(110)~\cite{krause_2009,bessarab_2013}. Therefore, assuming $\nu$ to have some fixed value independent of the system annihilation mechanism will lead to incorrect results, since the entropic and dynamical contributions are then not accounted for correctly.
Based on Monte-Carlo simulations, Hagemeister {\it et al.} \cite{hagemeister_2015} concluded that the attempt frequency for skyrmion annihilation was orders of magnitude smaller than that for skyrmion creation due to a broader shape of the skyrmion state energy minimum. An unusually small magnitude of the attempt 
frequency for thermally-activated skyrmion annihilation, on the order of $10^9$--$10^{10}$~s$^{-1}$, was confirmed by R\'ozsa {\it et al.}~\cite{rozsa_2016} and Rohart {\it et al.}~\cite{rohart_2016} who studied the skyrmion stability using Langevin spin dynamics simulations
at relatively high temperatures.

In the present work, we predict the lifetime of skyrmions in the Pd/Fe/Ir(111) system 
by evaluating the pre-exponential factor $\nu$ as well as the activation energy $\Delta E$ using harmonic transition state theory for spins~\cite{bessarab_2012} and  an atomistic spin Hamiltonian parametrized from DFT calculations. We show that the pre-exponential factor acquires a temperature dependence due to the presence of Goldstone modes, degrees of freedom for which the energy is constant. We find that the lifetime of an isolated skyrmion in a magnetic strip is  governed by at least two  annihilation mechanisms: Escape through the boundary and radial collapse at the interior.  We identify the crossover from one mechanism to the other as a function of applied magnetic field and temperature. While the calculations were performed for the Pd/Fe/Ir(111) system, the cross-over effect we predict should be a general feature for skyrmions in finite-size geometries and needs to be taken into account when designing novel logic or memory devices based on racetrack skyrmions.
 

\section{Results}

\noindent
\textbf{Skyrmion annihilation mechanisms in a racetrack.} 
Fig.~1 shows results of calculations of skyrmion annihilation in a Pd/Fe/Ir(111) racetrack geometry. Minimum energy paths (MEPs) between local energy minima corresponding to the skyrmion state and the field-polarized, ferromagnetic configuration are shown for different applied magnetic fields
(see the Methods section).
The calculations were carried out for magnetic fields at which the the field-polarized ferromagnetic configuration is the ground state, but individual skyrmions exist as metastable long-lived quasiparticles. 
For fcc-Pd/Fe/Ir(111), metastable isolated skyrmions within the saturated state are realized above a critical field $B_{\rm F}^{\text{fcc}}\approx3.2$~T at which the skyrmion state and the ferromagnetic state have the same energy. This field value is consistent with the measurements~\cite{romming_2015,kubetzka_2017}, although exact phase boundaries are not yet investigated experimentally. For hcp-Pd/Fe/Ir(111), the ferromagnetic phase is the ground state of the system over the whole range of magnetic field values, but skyrmions are metastable when the field is larger than $B_{\rm F}^{\text{hcp}}\approx 0.5$~T (see Ref.~\cite{malottki_2017}). 
The MEP calculations revealed two transition mechanisms, which are the same for both stackings of the Pd layer. The mechanisms are shown in Fig.~\ref{fig:1} for the fcc-Pd/Fe strip on Ir(111) (results for the hcp-Pd/Fe strip are presented in the Supplementary information). The first mechanism is characterized by a radial collapse of the skyrmion at the interior of the strip. It involves a symmetrical rotation of spins causing the skyrmion to gradually shrink and eventually disappear~\cite{bessarab_2015,lobanov_2016} (see Fig.~1(a,c)). The energy maximum along the MEP corresponds to a Bloch point-like texture, where the three central spins point opposite to each other (see snapshot \encircle{4} in Fig.~\ref{fig:1}). 

\begin{figure*}
\includegraphics[width=\textwidth]{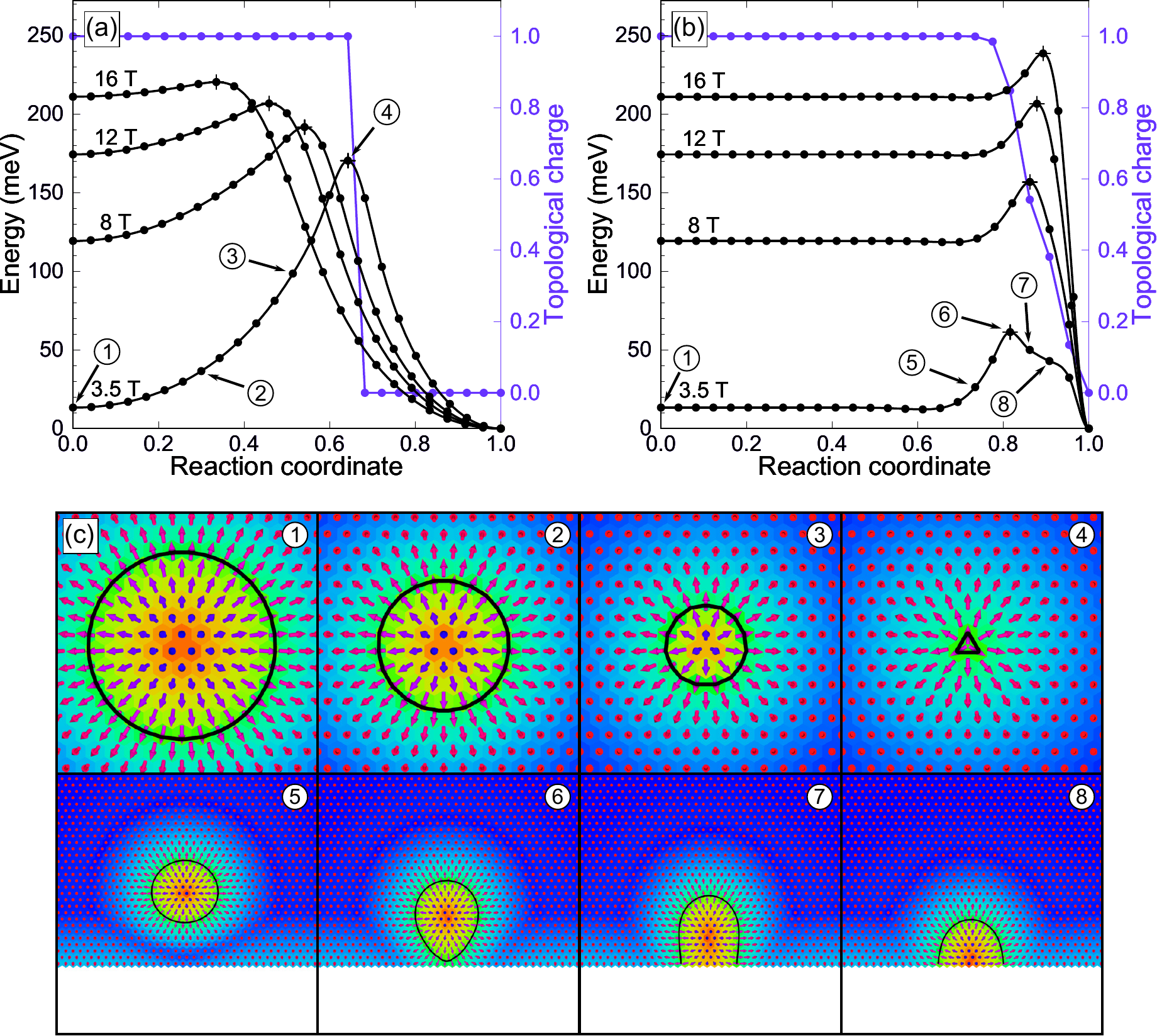}
\caption{{\bf Mechanisms of the skyrmion annihilation in a fcc-Pd/Fe/Ir(111) racetrack.} Energy variation along the MEPs for radial collapse of the skyrmion at the interior of the strip (a) and escape of the skyrmion through the boundary (b), shown for four values of applied magnetic field. The filled circles show position of the intermediate states along the annihilation paths, while crosses indicate energy maxima along the MEPs. Variation of the absolute value of the topological charge along the MEPs is represented by a purple line for $B=3.5$~T. The reaction coordinate is defined as the normalized displacement along the MEP. Starting- and end-point of the reaction coordinate is the skyrmion and ferromagnetic state, respectively.  The encircled numbers label the states for which spin configurations are shown in the lower panel (c). 
The background color indicates the value of the out-of-plane component of the magnetic vectors (red $\leftrightarrow$ up, blue $\leftrightarrow$ down). Black solid lines show the contour where the out-of-plane component of magnetization vanishes.
}
\label{fig:1}
\end{figure*}

The second mechanism corresponds to the skyrmion escaping through the boundary~\cite{stosic_2017,uzdin_2017,cortes_2017} (see Fig.~\ref{fig:1}(b,c)).
In the first section of the MEP,
the skyrmion moves as a whole without changing its size and shape towards the boundary of the strip. This translational motion of the skyrmion involves almost no change in energy. Repulsive interaction~\cite{rohart_2013} with the twisted moments of the under-coordinated boundary sites causes the skyrmion to deform as it approaches the boundary and the energy to rise (see snapshot \encircle{5} in Fig.~\ref{fig:1}).
At the energy maximum,
the skyrmion touches the edge of the strip, locally untwisting the spins at the boundary (see snapshot \encircle{6} in Fig.~\ref{fig:1}). 
The skyrmion then forms an excitation at the edge of the strip, which resembles the tail of a spin spiral. 
It
then leaves the sample by 
shrinking
resulting in a
linear decrease in energy along the MEP (see snapshots \encircle{7} and \encircle{8} in Fig.~\ref{fig:1}). Our calculations show that both mechanisms of skyrmion annihilation are realized in the system for the whole field range above $B_{\rm F}$, where skyrmions exist as metastable states.

Fig.~\ref{fig:1} 
also shows
the evolution of the absolute value of the topological charge $Q$ along the MEPs. For the collapse mechanism, $Q$ is unity until the saddle point is reached. At this point, it drops to zero. For the escape mechanism, the drop of the topological charge 
is not as abrupt and an indication of an inflection point is seen at $Q\approx 0.5$ where a meron is formed~\cite{alfaro_1976}. Under certain conditions, such excitations may constitute a long-lived, metastable states bound to the edge of the system~\cite{pereiro_2014}. 

\begin{figure*}
\includegraphics[width=\textwidth]{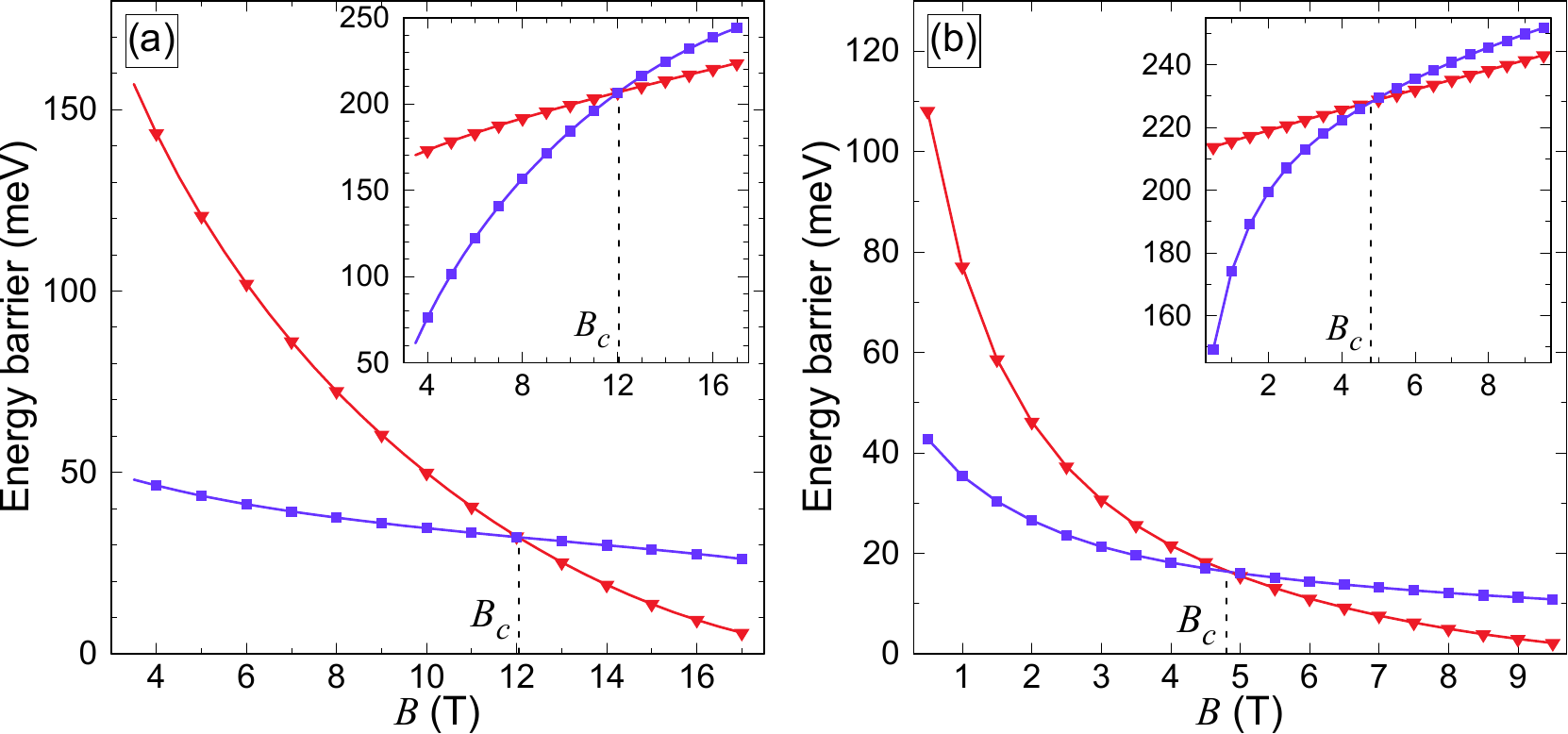}
\caption{{\bf Energy barriers for skyrmion annihilation in a Pd/Fe/Ir(111) racetrack.}
Energy barrier for the skyrmion annihilation and nucleation (inset) at the interior (red curve, triangles) and at the boundary (purple curve, squares) of the Pd/Fe strip as a function of applied magnetic field strength, shown for the fcc (a) and hcp (b) stacking of the Pd layer. The curves intersect at the crossover field, $B_c$.
}
\label{fig:2}
\end{figure*}

\vspace{\baselineskip}\noindent
\textbf{Energy barriers.}
 Noteworthy,  Fig.~~\ref{fig:1} exhibits that the energy barriers corresponding to the two pathways are not the same and they depend differently on the applied magnetic field.  The external magnetic field dependence of the energy barriers is shown in Fig.~\ref{fig:2} for the two annihilation mechanisms. At external fields at which single skyrmions emerge, $B=B_{\rm F}$, the energy required for the skyrmion to escape through the boundary is about one third of that for it to collapse in the interior
of the strip  for both fcc and hcp stacking of the Pd layer. With increasing external fields, barriers for both mechanisms decrease monotonically. However, the dependence is less pronounced for the escape mechanism. As a consequence, a crossover field, $B_c$, exists, above which the lower energy barrier is provided by the collapse at the interior of the strip. Clearly, the magnitude of $B_c$ depends on the value of the interaction parameters. For fcc-Pd/Fe/Ir(111), the crossover occurs at $B_c^{\text{fcc}}=12$~T, while calculations for hcp-Pd/Fe/Ir(111) give $4.8$~T for $B_c^{\text{hcp}}$. However, the presence of such a crossover between two mechanisms appears to be an inherent feature of skyrmions in confined geometries, and can be quite different in absolute values for different systems.

The determination of the energy barrier requires a very good knowledge of the entire energy landscape. In fact, we included DFT-derived exchange parameters for all atom pairs within 9 and 5 shells around each site for the fcc and hcp system, respectively (for more details see the First Principles Model Hamiltonian section.) For comparison, we also performed calculations using the effective parameters deduced from the experimental data~\cite{romming_2015} on field-dependent skyrmion profiles. In this case, the nearest-neighbor pairwise interactions were sufficient to model only a neighborhood of the local minimum  of the energy landscape corresponding to the metastable skrymion. Taking these effective parameters, then, the energy barriers are systematically smaller than  those obtained with DFT-calculated interaction parameters, but the crossover still persists, with $B_c^{\text{eff}}=2.6$~T (see Supplementary information). 

\vspace{\baselineskip}\noindent
\textbf{Skyrmion lifetimes.}
In order to verify that the crossover 
deduced from the energy barriers truly represents a
crossover between the annihilation mechanisms, the corresponding lifetimes and, therefore, the attempt frequency, $\nu$, need to be evaluated. 
Within harmonic transition state theory~\cite{bessarab_2012}, the Arrhenius pre-exponential factor 
is calculated based on the quadratic approximation of the energy surface at the transition state saddle point and the skyrmion state minimum. The harmonic approximation breaks down for possible Goldstone modes, along which the energy of the system does not change, and a special treatment of such modes is needed. 
Due to relatively large width of the strip, there are two Goldstone modes at the skyrmion state minimum corresponding to a translational motion in the plane of the film. The volume associated with this motion is proportional to the size of the track. There are no Goldstone modes at the saddle point corresponding to skyrmion collapse, but there are several such saddle points in the system, each contributing to the annihilation rate. In particular, the number of saddle points equals the number of interstitial sites, which also scales with the system size. 
As a result, the pre-exponential factor for the collapse mechanism does not depend on the size of the system.
The presence of Goldstone modes, however, introduces temperature dependence in the pre-exponential factor.
Each Goldstone mode at the 
initial
state contributes a factor of $\sqrt{2\pi k_B T}$ to the pre-exponential factor, while each Goldstone mode at the transition state gives a factor of $1/\sqrt{2\pi k_B T}$  (see the Methods section).  The prefactor for the collapse mechanism therefore scales with temperature as $\nu_{\rm col}(T)\sim T$. The saddle point for the boundary escape has one Goldstone mode corresponding to the translational motion of the excitation along the strip edge. As a result, the pre-exponential factor for the escape mechanism is inversely proportional to the width of the strip and scales with the square root of temperature, $\nu_{\rm esc}(T)\sim \sqrt{T}/W$. The results of the pre-exponential factor calculations for the 23.5~nm wide strip, several field strengths above the critical field for the low-temperature regime of 10~K are summarized in Table~\ref{tab:prefactor}. 

\begin{figure*}
\includegraphics[width=\textwidth]{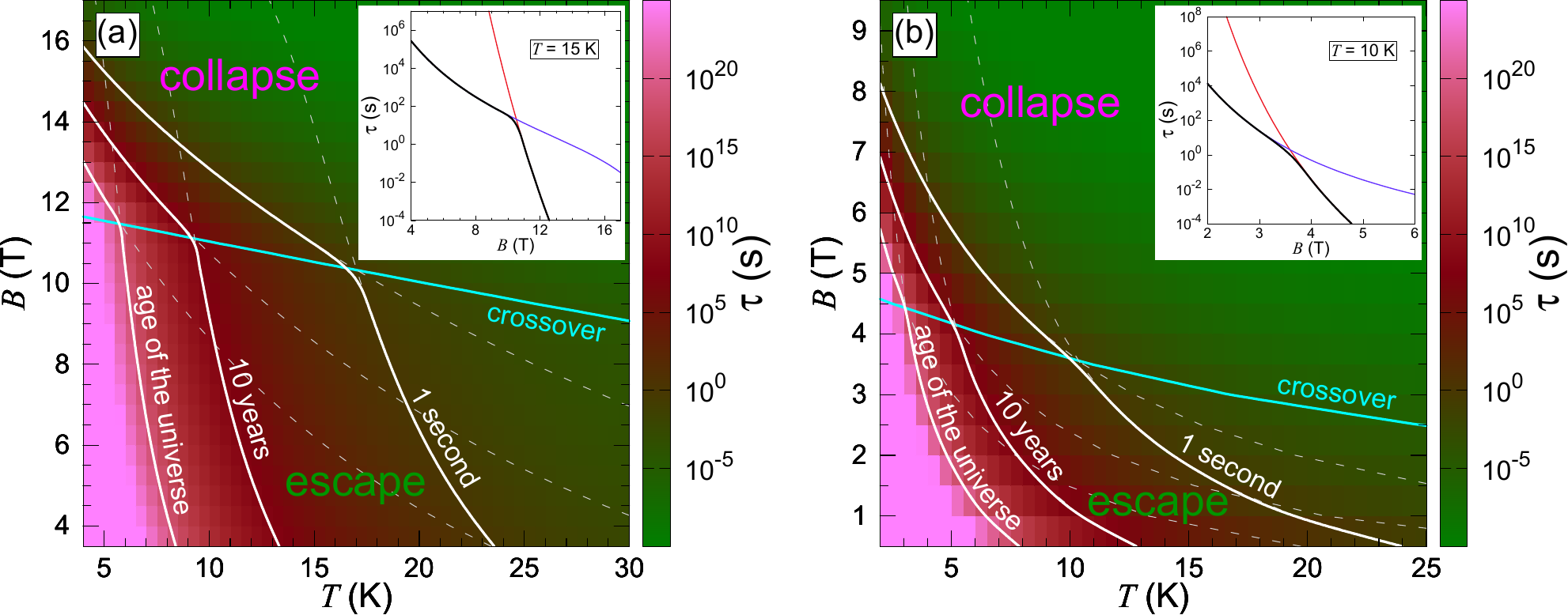}
\caption{{\bf Lifetime of a skyrmion in a Pd/Fe/Ir(111) racetrack.} Contour plot of the calculated lifetime of an isolated skyrmion in a 23.5~nm wide strip as a function of applied magnetic field strength and temperature, shown for the fcc (a) and hcp (b) stacking of the Pd layer. White contour lines have a characteristic cusp due to the crossover between collapse and escape mechanism indicated by the cyan line. White dashed lines indicate isochronal contours of the collapse and escape 
lifetimes.
Inset shows the cut of the contour plot at $T=15$ K (a) and $T=10$ K (b); in the insets, annihilation time due to collapse at the interior, escape through the boundary and total skyrmion lifetime are shown with red, purple and black curves, respectively.
}
\label{fig:3}
\end{figure*}

Relaxation time associated with each annihilation mechanism,
$\tau_{\rm col}$ and $\tau_{\rm esc}$, 
can now be calculated using Eq.~(\ref{eq:lifetime}). In addition to a strong, exponential dependence on temperature, both $\tau_{\rm col}$ and $\tau_{\rm esc}$ are characterized by a weak power dependence on temperature for the pre-exponential factor.
Moreover, $\tau_{\rm esc}$ scales with the strip width. This explicitly demonstrates that boundaries are less important for the stability of skyrmions in wider strips, as expected.  In Fig.~\ref{fig:3} the calculated results of the dependence of the skyrmion lifetime on the applied magnetic field and temperature are presented for a 23.5~nm wide strip, which is roughly five times larger than the skyrmion size at the critical field, $B_{\rm F}$. Both annihilation mechanisms contribute to the skyrmion stability, and the lifetime, $\tau$, is related to $\tau_{\rm col}$ and $\tau_{\rm esc}$ according to
\begin{equation}
    1/\tau=1/\tau_{\rm col}+1/\tau_{\rm esc},
\label{eq:lifetime_res}
\end{equation}
which follows from a general theory of two uncorrelated
processes.
Overall, for a given external field the contour graph exhibits an exponential dependence of the skyrmion lifetime on the temperature, and the lifetime changes at the critical field, $B_{\rm F}$, from the age of the universe to microsecond in a narrow temperature range of 25~K, which is about 10\% of the critical temperature for Pd/Fe/Ir system~\cite{boettcher_2017}. Two regions can be distinguished, corresponding to the two mechanisms of skyrmion annihilation. In the `collapse' region, $\tau_{\rm col}<\tau_{\rm esc}$, the skyrmion annihilation is dominated by the collapse mechanism, while in the `escape' region the lifetime is mostly defined by $\tau_{\rm esc}$. The two regions are separated by the crossover line defined as $\tau_{\rm col}=\tau_{\rm esc}=2\tau$. Notice, the crossover line is not parallel to the temperature axis, indicating that the crossover field is temperature dependent. At low temperature, the crossover field coincides with that of the lowest energy barrier. At higher temperature, entropy comes into play, leading to a decrease in the crossover field. Although fcc and hcp stackings of the Pd layer result in different skyrmion lifetime, the results for both stackings share the same characteristic features (also see the results for the effective, nearest neighbor Hamiltonian from Ref.~\cite{romming_2015} in the supplementary information).

\section*{Discussion}
Our results show that boundaries have a crucial impact on the skyrmion stability in 
racetrack strips at low magnetic fields. If the system boundaries are not included in the analysis, the skyrmion lifetime in the Pd/Fe/Ir(111) system at the field $B=B_{\rm F}$, at which metastable skyrmions emerge at a ferromagnetic background, is expected to be ten years at $T\approx$ 30--35~K for both fcc and hcp stacking of the Pd layer (see dashed contour lines of $\tau_{\rm col}$ in Fig.~\ref{fig:3}). 
This, by the way, is in significant difference to results deduced from Monte Carlo simulations fitted to experimental measurements of the skyrmion solution in an extended Pd/Fe/Ir(111) film,~\cite{hagemeister_2015}
indicating that 19~K is needed to achieve mean lifetimes on the order of years, a temperature at which we achieve lifetimes in the order of the lifetime of the universe. 

However, considering a magnetic strip instead of a two-dimensional film, a ten-year lifetime, which is sufficient for information storage, is achieved at $B=B_{\rm F}$ only below 15~K in the strip of width, $W$, five times larger than the skyrmion size, $L_{\rm s}$, at $B_{\rm F}$, $W\simeq 5\,L_{\rm s}$. Interestingly the lifetime for the escape process is proportional to the strip width, $\tau_{\rm esc}\propto W$, which is a consequence of smaller probability for a skyrmion to reach the boundary in a wider strip. Thus, the above  temperature can be raised to 30~K by increasing the strip width. A rough estimate requires strip widths wider by a factor larger than 10$^8$, which are totally unrealistic in practise.
Clearly, for any practical purpose, the escape mechanism is the most important one for skyrmion annihilation and nucleation in the Pd/Fe/Ir systems. 
In order to enhance the skyrmion stability at magnetic fields at which metastable skyrmions emerge, decoration of some kind at the boundaries of the strip is needed.

At larger applied fields, the collapse mechanism becomes increasingly important and completely dominates the skyrmion annihilation process above the crossover field. The crossover between the annihilation mechanisms is to a large extent due to the crossover of the lowest energy barrier (compare insets in Fig.~\ref{fig:3} and Fig.~\ref{fig:2}), which has a simple interpretation.
Consider the effect of skyrmion size, $L_s$. 
The energy barrier associated with the collapse mechanism is proportional to the number of spins 
that need to be
flipped in this process and, therefore, scales with $L_{\rm s}^2$. As expected, the barrier goes to zero as the skyrmion size approaches the size of the saddle point excitation, which essentially represents a Bloch point-like defect. The size of the defect is small but finite and it appears to be roughly field independent. On the other hand, the energy barrier for the escape mechanism is defined by the number of spins 
that need to be
untwisted at the boundary, not by the shrinking of the skyrmion.
When the skyrmion leaves the system, it untwists the boundary locally, with the number of spins involved 
scaling as
$\sim L_{\rm s}$. Since the skyrmion size is a monotonous function of the applied magnetic field~\cite{Bogdanov1989,Bogdanov1994,romming_2015}, the weaker dependence of the escape barrier on $L_{\rm s}$ results in a weaker dependence on $B$. On the other hand, the energy barrier for the escape mechanism is nonzero even at an upper critical field, $B_u$, at which the skyrmion solution becomes unstable and the barrier for the collapse mechanism vanishes, because the skyrmion size is always finite and the boundary twist exists even at very large fields, much larger than $B_u$~\cite{meynell_2014,leonov_2016}. This explains why curves $\Delta E_{\rm col}(B)$ and $\Delta E_{\rm esc}(B)$ must intersect. This conclusion is rather general and should be valid for any finite-size skyrmion system.

It is interesting to compare the barrier for the skyrmion creation and annihilation with the well-known energy of a Bloch point singularity of topological charge $|Q|=1$~\cite{belavin_1975,tretiakov_2007}, through which the skyrmion can collapse within the micromagnetic continuum model. This energy serves as a standard scale for barriers. With the exchange energy described as $E(\vec{m})=\int_{\mathbb{R}^2} A |\vec{\nabla}\vec{m}|^2 d\vec{r} $ in a two-dimensional model, the Bloch point energy with respect to the ferromagnetic state amounts to $E_{\text{BP}}=8\pi A$~\cite{tretiakov_2007}. Taking an experimentally deduced value for the exchange stiffness $A$~\cite{romming_2015}, we obtain a nucleation barrier of 128~meV, which is about 60\% larger compared to the barrier height calculated within the atomistic spin Hamiltonian equipped with the equivalent effective, nearest-neighbor exchange interaction parameter, $J_{\text{eff}} = 2A/\sqrt{3}$ (see the inset in Supplementary Fig. 2). The agreement between the models is surprisingly good, given that the micromagnetic model completely ignores the structure of the transition state on the atomic lattice scale. 
Noteworthy, the continuum-model estimate approaches the values obtained with the shell-resolved exchange Hamiltonian for the fcc-Pd/Fe/Ir(111) system (see the inset in Figs.~\ref{fig:2}(a)), which, however, is likely a coincidence. We therefore conclude that the continuum-model estimate can be used to predict the characteristic scale for the barrier, but is insufficient for the quantitative determination of the skyrmion stability and lifetime.

In this paper we focused on the long-lifetime limit as relevant for technology associated with the low-temperature regime for the Pd/Fe/Ir(111) systems. Here, we briefly comment on the short lifetime regimes. Typical magnetic time scales are microsecond (MHz), nanosecond (GHz) and picosecond (THz) regimes, related to skyrmion core gyration dynamics, magnon excitation and spin precession, respectively. Staying at external magnetic fields $B_{\rm F}$, where single skyrmions are stable at the ferromagnetic background, and large sample sizes, where the escape mechanism becomes irrelevant, the lifetime of the single skyrmion matches these timescales at around 80~K, 140~K, and 200~K, respectively. Thus, in case of interest in the  fabrication of frequency-tunable spin-torque vortex oscillators, the operation temperature should be significantly below 80~K. When skyrmion annihilation and creation interferes with magnon excitations, also the excitation of short-lived antiskyrmions or skyrmions of different charges are possible. This is consistent with the intermediate phase regime of skyrmions and antiskyrmions introduced in Ref.~\cite{boettcher_2017}. The spin-precession time coincides with the lifetime of skyrmions around the critical temperature of the ferromagnetic phase to the paramagnetic phase. In that situation, of course the lifetime is not anymore a relevant concept.

We believe that the theoretical approach used here, which combines a statistical description of skyrmion stability  with an atomistic spin Hamiltonian parametrized from first-principles electronic structure calculations gives valuable insight into the main mechanisms governing the lifetime of skyrmions and provides a tool for predictive simulation-based materials design for skyrmion-based information technology. While the focus here was on the high-field, low-temperature, Pd/Fe/Ir(111) model-system based nanotrack, the exponential temperature dependence, the importance of the finite width on the life-time at small fields and the crossover effect at higher fields combined with a finite temperature range for technologically long skyrmion life times should be general features for skyrmions in finite-size systems, including multilayer heterostructures, where skyrmions exist at room temperature~\cite{moreau_2016,boulle_2016,woo_2016}.


\section*{Methods}

\noindent
\textbf{Simulated system.}
We model the Pd/Fe film as a single monolayer of classical spins on a hexagonal lattice 
defined by the Ir(111) substrate. The lattice constant is 2.7~\AA~\cite{dupe_2014}.
While the Fe layer follows the fcc stacking of the Ir(111) surface, both 
fcc and hcp stackings for the Pd atoms
have been observed experimentally~\cite{kubetzka_2017}
and DFT calculations have been carried out for both types of layers, resulting in 
two sets of magnetic interaction parameters~\cite{malottki_2017}. 
The two structures are referred to as fcc-Pd/Fe/Ir(111) and hcp-Pd/Fe/Ir(111), respectively.

A magnetic racetrack is simulated by applying periodic boundary conditions along one of the lattice-bond directions, while imposing open boundary conditions along the orthogonal direction. The size of the computational domain was chosen to be large enough for the isolated skyrmion at the center of the strip not to be affected by the boundaries. 
The width of the strip is taken to be $W=23.5$~nm.

\vspace{\baselineskip}\noindent
\textbf{First-Principles Model Hamiltonian.} 
The energy landscape of the Pd/Fe monolayer strip on Ir(111) is described by the following atomistic Hamiltonian:
\begin{eqnarray}
E = - \sum_{i,j}
J_{ij}^{X}\vec{m}_i  \cdot  \vec{m}_j - 
\sum_{i,j} 
\vec{D}_{ij}^{X}  \cdot  [\vec{m}_i \times  \vec{m}_j]
\nonumber\\
-K^{X} \sum_{i} \left(\vec{m}_i \cdot \hat{e}_K\right)^2 
- \mu_\mathrm{s} \vec{B}\sum_{i}\vec{m}_i.         
\label{eq:ham}
\end{eqnarray}
Here, summation in the first two terms runs over 
distinct pairs of atoms,
$\vec{m}_i$ is a unit vector defining the orientation of the magnetic moment at site $i$. 
The superscript $X$ indicates the parameter set: $X=\text{fcc}$ for the fcc-Pd/Fe/Ir(111) system, while $X=\text{hcp}$ for the hcp-Pd/Fe/Ir(111) one. The DM vector, $\vec{D}_{ij}$, is a three-dimensional vector with components in the monolayer plane that point perpendicular to the bond connecting sites $i$ and $j$ and small positive and negative components normal to the surface. Since these normal components average out after summation over all pairs, they are neglected here. The unit vector $\hat{e}_K$, defining the easy axis as well as the external magnetic field $\vec{B}$, point perpendicular to the film plane. Shell-resolved exchange interaction parameters, $J_{ij}$, DM interaction parameters, $D_{ij}$, anisotropy constant, $K$ as well as on-site magnetic moment, $\mu_s$, obtained from first-principles calculations were taken from Ref.~\cite{malottki_2017}. In order to describe the energy of the states along the MEP we included the interaction parameters due to exchange, $J_{ij}$, for the pairwise contributions including up to 9, 5 shells with in total 72, 36 pairs per atom for the fcc, hcp stacked Pd/Fe/Ir system, respectively. 

\vspace{\baselineskip}\noindent
\textbf{Evaluation of annihilation rates.}
The rate of skyrmion annihilation, which defines the lifetime, is calculated using harmonic transition state theory for magnetic systems~\cite{bessarab_2012} 
extended to include the presence of Goldstone modes.
The theory predicts an Arrhenius dependence of the rate on temperature (see Eq.~(\ref{eq:lifetime}) in the main text), where the activation energy is given by the energy difference between the relevant saddle point and a local minimum on the energy surface corresponding to the initial state, while the pre-exponential factor is defined by the curvature of the energy surface at the saddle point and at the initial state minimum. If no Goldstone modes are present in the system, the Boltzmann average entering the equation for the transition rate is computed using Gaussian integration of the distribution function, $\rho=C\exp\left[-E/(k_\text{B} T)\right]$, over all relevant degrees of freedom, resulting in the factor $\sim (2 \pi k_\text{B} T)^{-N}$ for a system of $N$ spins. The same factor enters the normalization constant, $C\sim(2 \pi k_\text{B} T)^{N}$. As a result, the temperature dependence cancels out in the expression for the attempt frequency~\cite{bessarab_2012}. The harmonic approximation breaks down for possible Goldstone modes appearing in the system. Specifically, integration of the partition function over a Goldstone mode gives the volume in the phase space associated with the mode. If the numbers of Goldstone modes are not the same at the minimum and at the saddle point, some of  the factors $\sqrt{2 \pi k_\text{B}}$ resulting from the Gaussian integration remain uncompensated and the pre-exponential factor acquires a power dependence on temperature. Specifically, the pre-exponential factor is given by the following equation, generalized to include Goldstone modes:
\begin{equation}
\label{eq:prefactor0}
\nu = \frac{\gamma}{2\pi}\frac{V_\text{SP}}{V_\text{min}}\left(2\pi k_\text{B} T\right)^{(P_\text{min}-P_\text{SP})/2}
\sqrt{ \sideset{}{^\prime}\sum_i \frac{a_i^2}{\epsilon_{i}}}
\sqrt{\frac{\det \mathbf{H}_\text{min}}{\det^{\prime} \mathbf{H}_\text{SP}}}.
\end{equation}
Here, 
$\gamma$ is the
gyromagnetic ratio, $\det \mathbf{H}_\kappa$, $P_\kappa$  and $V_\kappa$ denote the product of the eigenvalues of the Hessian matrices $\mathbf{H}_\kappa$, numbers of Goldstone modes and volumes associated with the Goldstone modes, respectively, at the stable state minimum ($\kappa=\text{min}$) and at the saddle point ($\kappa = \text{SP}$), index $i=1,\ldots,2N$ labels degrees of freedom in the system, $\epsilon_i$ are the eigenvalues of the Hessian at the saddle point, and $a_i$ are expansion coefficients in the linearized equation for the unstable mode derived from the Landau-Lifshitz equations of motion for $2N$ degrees of freedom of the system~\cite{bessarab_2012}. 
The terms corresponding to the Goldstone modes are omitted in the determinants and the summation. The terms associated with the unstable mode are omitted as well, as indicated by the prime superscript.

\vspace{\baselineskip}\noindent
\textbf{Minimum energy path calculations.}
Calculation of minimum energy paths (MEPs) is needed for a definitive identification of transition state saddle points, which define the transition rates within the harmonic transition state theory.  A MEP between two minima is the path in configuration space which lies lowermost on the energy surface. Following a MEP means rotating spins in an optimal way, so that the energy is minimal with respect to all degrees of freedom perpendicular to the path. The MEP not only gives the location of the saddle point, which corresponds to a maximum along the MEP, but also provides detailed information about the transition mechanism, important quantitative knowledge, which is not easily accessible in the spin dynamics simulations. A Geodesic Nudged Elastic Band (GNEB) method~\cite{bessarab_2015,bessarab_2017} is used to find MEPs of skyrmion annihilation. The GNEB method involves taking some initial guesses of the path represented by a discrete chain of states of the system, and systematically bringing that to the nearest MEP by zeroing the transverse component of the gradient force at each point along the path described in the following. In order to distribute the states evenly along the path, virtual springs are introduced between adjacent states. At each state, a local tangent to the path is estimated and the GNEB force guiding the states towards the MEP is defined as the sum of the transverse component of the negative energy gradient and the component of the spring force along the tangent. The positions of states are then adjusted using some optimization algorithm so as to zero the GNEB force. In the method, both the GNEB force and the path tangent are defined in the local tangent space to the curved configuration space, which is needed to satisfy constraints on the length of magnetic moments and to properly decouple the perpendicular component of the energy gradient from the spring force~\cite{bessarab_2015}.
%


\section{Acknowledgments}
The authors would like to thank S. Heinze, S. von Malottki, B. Dup\'e, D. Thonig, B. Hj\"orvarsson, A. Bergman, M. Pereiro, O.A. Tretiakov for helpful discussions and useful comments. We acknowledge financial support from the Icelandic Research Fund (Grant No. 163048-052), the mega-grant of the Ministry of Education and Science of the Russian Federation (grant no. 14.Y26.31.0015), G\"oran Gustafsson Foundation, Vetenskapsr\r{a}det (VR), Carl Tryggers Stiftelse (CTS), the European Union’s Horizon 2020 research and innovation programme (grant agreement no.\ 665095 -- FET-Open project MAGicSky),
Academy of Finland (grant no. 278260), 
and Swedish Energy Agency (STEM). P.F.B. acknowledges support from the Russian Science Foundation (Grant No. 17-72-10195). Some of the computations were performed on resources provided by the Swedish National Infrastructure for Computing (SNIC) at the National Supercomputer Center (NSC), Link\"oping University, the PDC Centre for High Performance Computing (PDC-HPC), KTH, and the High Performance Computing Center North  (HPC2N), Ume\r{a} University.
\section{Author contributions}
P.F.B., A.D., G.P.M., F.N.R and N.S.K. conceived the project. P.F.B. performed the calculations. P.F.B, A.D., H.J., and S.B. wrote the manuscript. 
I.S.L., H.J. and V.M.U. contributed to the prefactor analysis. 
All authors analyzed and discussed the data and contributed to the writing of the article.
\section{Competing financial interests}
The authors declare that they have no
competing financial interests.
\section{Correspondence}
Correspondence and requests for materials
should be addressed to A.D.~(email: annadel@kth.se) or P.F.B.~(email: bessarab@hi.is).

\begin{table*}
\caption{{\bf Pre-exponential factors for skyrmion annihilation in a Pd/Fe/Ir(111) racetrack.} The pre-exponential factors for the skyrmion collapse, $\nu_{\rm col}$, and escape, $\nu_{\rm esc}$, in a 23.5 nm wide Pd/Fe strip on Ir(111) for several applied magnetic field strengths and a temperature of 10 K. 
Calculations have been carried out for both
fcc and hcp stackings of the Pd layer
}
\begin{center}
\begin{tabularx}{0.9\textwidth}{YYY|YYY}
\hline\hline
\multicolumn{3}{c|}{fcc-Pd/Fe/Ir(111)} & \multicolumn{3}{c}{hcp-Pd/Fe/Ir(111)} \\
$B$ (T) & $\nu_{\rm col}$ (s$^{-1}$) & $\nu_{esc}$ (s$^{-1}$) & $B$ (T) &  $\nu_{\rm col}$ (s$^{-1}$) & $\nu_{esc}$ (s$^{-1}$) \\
\hline
4 & $4.0\cdot 10^{14}$ & $1.2\cdot 10^{10}$ & 1 & $2.7\cdot 10^{10}$ & $1.1\cdot 10^{9}$ \\
5 & $2.7\cdot 10^{14}$ & $1.1\cdot 10^{10}$ & 2 & $4.4\cdot 10^{11}$ & $1.8\cdot 10^{9}$ \\
6 & $1.8\cdot 10^{14}$ & $1.1\cdot 10^{10}$ & 3 & $1.2\cdot 10^{12}$ & $2.4\cdot 10^{9}$ \\
7 & $1.2\cdot 10^{14}$ & $1.1\cdot 10^{10}$ & 4 & $2.0\cdot 10^{12}$ & $2.9\cdot 10^{9}$ \\
8 & $8.4\cdot 10^{13}$ & $1.1\cdot 10^{10}$ & 5 & $2.5\cdot 10^{12}$ & $3.3\cdot 10^{9}$ \\
9 & $5.9\cdot 10^{13}$ & $1.0\cdot 10^{10}$ & 6 & $2.8\cdot 10^{12}$ & $3.7\cdot 10^{9}$ \\
10 & $4.3\cdot 10^{13}$ & $1.0\cdot 10^{10}$ & 7 & $2.9\cdot 10^{12}$ & $4.0\cdot 10^{9}$ \\
11 & $3.1\cdot 10^{13}$ & $1.0\cdot 10^{10}$ & 8 & $3.0\cdot 10^{12}$ & $4.3\cdot 10^{9}$ \\
12 & $2.3\cdot 10^{13}$ & $1.0\cdot 10^{10}$ & 9 & $3.0\cdot 10^{12}$ & $4.7\cdot 10^{9}$ \\
13 & $1.7\cdot 10^{13}$ & $1.0\cdot 10^{10}$ & - & - & - \\
14 & $1.3\cdot 10^{13}$ & $1.0\cdot 10^{10}$ & - & - & - \\
15 & $9.7\cdot 10^{12}$ & $1.1\cdot 10^{10}$ & - & - & - \\
16 & $7.3\cdot 10^{12}$ & $1.2\cdot 10^{10}$ & - & - & - \\
17 & $5.5\cdot 10^{12}$ & $1.5\cdot 10^{10}$ & - & - & - \\
\hline\hline
\end{tabularx}
\end{center}
\label{tab:prefactor}
\end{table*}

\end{document}